\documentclass[showpacs,amssymb,preprint,preprintnumbers,nofootinbib,superscriptaddress]{revtex4}
\usepackage{epsf,epsfig,graphics,graphicx}
\begin{document}

%\title{Electromagnetic coupling effects in natural inflation}
%
%\author{Hing-Tong Cho$^1$}
%\author{Kin-Wang Ng$^{2,3}$}
%
%\affiliation{
%$^1$Department of Physics, Tamkang University, Tamsui, New Taipei City 25137, Taiwan\\
%$^2$Institute of Physics, Academia Sinica, Taipei 11529, Taiwan\\
%$^3$Institute of Astronomy and Astrophysics, Academia Sinica, Taipei 11529, Taiwan}

\title{Electromagnetic coupling effects in natural inflation}
\author{Hing-Tong Cho}
\email[Email: ]{htcho@mail.tku.edu.tw}
\affiliation{Department of Physics, Tamkang University, Tamsui, New Taipei City 25137, Taiwan}
\author{Kin-Wang Ng}
\email[Email: ]{nkw@sinica.phys.edu.tw}
\affiliation{Institute of Physics, Academia Sinica, Taipei 11529, Taiwan}
\affiliation{Institute of Astronomy and Astrophysics, Academia Sinica, Taipei 11529, Taiwan}

\date{\today}

\begin{abstract}
In this work we study the effects of the electromagnetic coupling in natural inflation in a systematic manner using the Schwinger-Keldysh formalism. The corresponding influence functional is evaluated to one-loop level. It can be interpreted as due to a single stochastic force. The equation of motion of the inflaton field is therefore given in the form of a Langevin equation. Lastly, the two-point and the three-point correlation functions of the inflaton field are worked out. They are related to the power spectrum and the nongaussianity of the inflaton field, respectively.
\end{abstract}

\pacs{98.80.Cq, 04.62.+v}
\maketitle

\section{Introduction}

In order to remedy some of the problems of the standard big bang theory, an inflationary period has been introduced in the very early universe before the radiation dominated era~\cite{guth}. For this inflation paradigm to work, the corresponding inflaton potential must be very flat, or that the inflaton has to roll slowly to have sufficient e-folding expansion and large-scale matter homogeneities. This constitutes a fine-tuning problem. A natural way to avoid this is to take the inflaton to be axion-like, which possesses a naturally flat potential with the shift symmetry~\cite{freese,adams}. This model of inflation is known as the natural inflation, followed by many of its variant models~\cite{axion}.

One major problem of the natural inflation is that the symmetry breaking scale has to be very close to the Planck scale to have a spectral index for the matter power spectrum compatible with observations~\cite{savage}. It was noted that this scale could be lowered to the GUT scale in a warm inflation scenario~\cite{mohanty,anber,visinelli,mishra}. In warm inflation models~\cite{berera1,berera2,wolung,berera3} the couplings of the inflaton field to other fields are considered during the inflationary period. This is like the coupling to a heat bath in which noise and dissipative effects will be manifest. Specifically the appearance of the dissipative term could prevent the inflaton field from rolling down the potential quickly even for a steep potential, thereby lowering the constraint on the energy scale.

Recently, a lot of attentions have been paid to the study of the phenomenological effects of the axion-vector coupling
on inflation~\cite{anber2,anber,durrer,barnaby,meer,ng,barnaby2,cook,lin,linde,bugaev,fugita,adshead,cheng,ferreira},
with particular attentions to the influences of particle production
and its associated backreaction to the slow-roll inflation.
This leads to very interesting effects such as the
generation of non-Gaussian and non-scale-invariant density power
spectrum~\cite{barnaby,meer}, the damping of primordial
gravitational waves due to a copious production of free-streaming particles~\cite{ng},
and the generation of stochastic gravitational waves that
could be detected at ground-based gravity-wave interferometers~\cite{barnaby2,cook},
and the formation of primordial black holes at density peaks near
the end of inflation~\cite{lin,linde,bugaev}.

Despite the above mentioned works, we find that there is lacking a systematic study of the origin of the effects
caused by the axion-vector coupling in the natural inflation.
In order to understand the effect of the heat bath in more details, one could utilize the tools of the closed time path (CTP)~\cite{jordan,calzetta}, or the Schwinger-Keldysh formalism~\cite{schwinger,keldysh}, together with the influence functional method~\cite{feynman}. In this open system approach the degrees of freedom of the heat bath are integrated out. This coarse graining process produces the noise and the dissipation terms~\cite{calzettahu} that the system is subjected to. In the warm natural inflation model, the axion-like inflaton is coupled to the electromagnetic field. One could therefore integrate out the photon degrees of freedom giving an influence functional with the corresponding noise and dissipation terms. In this paper we would explore this possibility, and then we would study the effects of these terms to the evolution of the inflaton field.

\section{CTP formalism}

To begin with we consider the following action of axion inflation with the inflaton field $\varphi (x)$ and its interaction with the photon field,
\begin{equation}
S_{tot}=\int d^{4} x\sqrt{-g}  \left[\frac{M_{P}^{2} }{{2}} R-\frac{1}{2} \left(\partial \varphi \right)^{2} -V(\varphi )-\frac{1}{4} F^{\mu \nu } F_{\mu \nu } -\frac{\alpha }{4f} \varphi \, \tilde{F}^{\mu \nu } F_{\mu \nu } \right],
\label{Stot}
\end{equation}
where $R$ is the curvature scalar, $M_{P}$ the reduced Planck mass, $V(\varphi)$ the inflaton potential,
$f$ the axion decay constant, and $\alpha $ is a dimensionless parameter. The simplest example of the inflaton potential would be
\begin{eqnarray}
V(\varphi)=\Lambda^{4}\left[1+\cos\left(\frac{\varphi}{f}\right)\right].
\end{eqnarray}
with the mass scale $\Lambda\sim m_{GUT}$ and $f<M_{P}$.

Here we take the spacetime to be conformally flat,
\begin{equation}
ds^{2} =a^{2} (\tau )(-d\tau ^{2} +d\vec{x}\cdot d\vec{x}),
\label{flatmetric}
\end{equation}
where $a(\tau)$ is the cosmic scale factor that depends on conformal time. The electromagnetic field is $F_{\mu \nu } =\partial _{\mu } A_{\nu } -\partial _{\nu } A_{\mu } $ and $\tilde{F}^{\mu\nu} = \frac{1}{2}\epsilon^{\mu\nu\alpha\beta} F_{\alpha\beta}/{\sqrt{-g}}$ is its dual, where the potential $A_{\mu }$ satisfies the gauge conditions: $A_{\tau} =0$ and $\nabla ^{\mu } A_{\mu } =0$. Then one can write $A_{\mu } =(0,\vec{A})$ and expand $\vec{A}$ in terms of the two physical degrees of freedom,
\begin{equation}
\vec{A} (\vec{x},\tau )=\sum _{\lambda =R,L}\int \frac{d^{3} k}{(2\pi )^{3/2} }  {\kern 1pt} \, e^{i\vec{k}\cdot \vec{x}} A_{\lambda } (\vec{k},\tau )\vec{\varepsilon }_{\lambda } (\vec{k}),
\end{equation}
where $\vec{\varepsilon }_{\lambda } (\vec{k})$ are the circular polarization vectors. They have the following properties,
\begin{eqnarray}
&&\vec{\varepsilon }_{R,L}^{*} (\vec{k})=\vec{\varepsilon }_{L,R} (\vec{k})=\vec{\varepsilon }_{R,L} (-\vec{k});\nonumber\\
&&\vec{\varepsilon }_{\lambda }^{*} (\vec{k})\cdot \vec{\varepsilon }_{\lambda'} (\vec{k})=\delta _{\lambda \lambda'},\ \ \
\vec{k}\cdot \vec{\varepsilon }_{\lambda } (\vec{k})=0;\nonumber\\
&&\vec{k}\times \vec{\varepsilon }_{R,L} (\vec{k})=\mp i|\vec{k}|\vec{\varepsilon }_{R,L} \,.
\label{polvec}
\end{eqnarray}

In terms of the physical modes the free photon action can be expressed as
\begin{eqnarray}
&&\int d^{4} x\sqrt{-g}  \left(-\frac{1}{4} F^{\mu \nu } F_{\mu \nu } \right)\nonumber\\
&&=\frac{1}{2} \int d\tau \sum _{\lambda }\int d^{3} k\left[\left(\partial _{\tau} A_{\lambda }^{*} (\vec{k},\tau )\right)\left(\partial _{\tau} A_{\lambda } (\vec{k},\tau )\right)-\vec{k}^{2} A_{\lambda }^{*} (\vec{k},\tau )A_{\lambda } (\vec{k},\tau )\right].
\end{eqnarray}
Note that these two free physical photon modes are equivalent to two independent massless scalar fields, defined by
\begin{equation}
A_{\lambda} (\vec{x},\tau )=\int \frac{d^{3} k}{(2\pi )^{3/2} }  {\kern 1pt} \, e^{i\vec{k}\cdot \vec{x}} A_{\lambda } (\vec{k},\tau ).
\label{scalar}
\end{equation}
Next, we consider the inflaton photon interaction term. We have
\begin{eqnarray}
&&-\int d^{4} x\sqrt{-g} \frac{\alpha }{4f}\, \varphi \, \tilde{F}^{\mu \nu } F_{\mu \nu }\nonumber \\
&&=-\frac{\alpha }{f} \int d^4 x\, \varphi (x)
\sum _{\lambda'}\int \frac{d^{3} k'}{(2\pi )^{3/2} }  {\kern 1pt} \, e^{i\vec{k}'\cdot \vec{x}} \left(\partial _{\tau} A_{\lambda' } (\vec{k}',\tau )\right)\left(\vec{\varepsilon }_{\lambda' } (\vec{k'})\right)_i\nonumber\\
&&\quad\times\int \frac{d^{3} k}{(2\pi )^{3/2} }  {\kern 1pt} \, e^{i\vec{k}\cdot \vec{x}}
|\vec{k}| \left[A_{R}(\vec{k},\tau ) \left(\vec{\varepsilon }_{R} (\vec{k})\right)_i
-A_{L} (\vec{k},\tau ) \left(\vec{\varepsilon }_{L} (\vec{k})\right)_i\right].
\end{eqnarray}

To proceed we look more closely at the interaction between the two physical modes of the photon and the inflaton. To do that we expand the inflaton field around a homogeneous background, $\varphi (\vec{x},\tau )=\Phi (\tau )+\phi (\vec{x},\tau )$. First we consider the effect of the homogeneous background on the photon modes. For the homogeneous term, the inflaton photon interaction term becomes
\begin{eqnarray}
&&-\int d^{4} x\sqrt{-g}  \frac{\alpha }{4f} \, \Phi \, \tilde{F}^{\mu \nu } F_{\mu \nu }\nonumber \\
&&=\frac{\alpha }{2f} \int d\tau \left(\partial _{\tau} \Phi (\tau )\right)\int d^{3} k|\vec{k}|\left[A_{R}^{*} (\vec{k},\tau )A_{R} (\vec{k},\tau )-A_{L}^{*} (\vec{k},\tau )A_{L} (\vec{k},\tau )\right], \
\end{eqnarray}
where integrations by parts with respect to $\tau $ have been performed. This part can be combined with the free photon action to give
\begin{equation}
-\frac{1}{2} \int d\tau \sum _{\lambda =R,L}\int d^{3} k{\kern 1pt} {\kern 1pt} {\kern 1pt} A_{\lambda }^{*} (\vec{k},\tau )\left[\partial _{\tau}^{2} +\vec{k}^{2} \mp 2aH\xi |\vec{k}|\right]   A_{\lambda } (\vec{k},\tau ),
\label{quadratic}
\end{equation}
where we have introduced the Hubble parameter and a dimensionless quantity, respectively,
\begin{equation}
H\equiv {1\over a}\frac{da}{dt}, \quad\quad  \xi\equiv \frac{\alpha}{2fH} \frac{d\Phi}{dt}.
\end{equation}

Due to the coupling with the homogeneous background of the inflaton field, $A_{R} (\vec{k},\tau )$ will grow with $\tau $, while $A_{L} (\vec{k},\tau )$ will be exponentially suppressed (see Appendix~\ref{mode}). In short, as long as the parameter $\xi\gg|k\tau|$, $A_{R}(\vec{k},\tau)$ will be amplified by a factor of $e^{\pi\xi}$, while the other polarization $A_{L}(\vec{k},\tau)$ will have no such amplification. In the following consideration we assume that the parameter $\xi$ is indeed large enough for this to happen. We will therefore retain only $A_{R} (\vec{k},\tau )$ and will treat it as a scalar field $A(\vec{x},\tau )$ with the above quadratic action~(\ref{quadratic}). Hence we need to deal with the interaction term with $\phi (\vec{x},\tau )$, which becomes
\begin{eqnarray}
&&-\int d^{4} x\sqrt{-g}  \frac{\alpha }{4f} \, \phi \, \tilde{F}^{\mu \nu } F_{\mu \nu } \nonumber\\
&&=-\frac{\alpha }{f} \int d^4 x\, \phi (x)
\int \frac{d^{3} k' d^{3} k}{(2\pi )^{3} }  {\kern 1pt} \, e^{i(\vec{k}'+\vec{k})\cdot \vec{x}} \left(\partial _{\tau} A (\vec{k}',\tau )\right)
A(\vec{k},\tau ) |\vec{k}|(\vec{\varepsilon }_{R} (\vec{k}')\cdot \vec{\varepsilon }_{R} (\vec{k})).
\end{eqnarray}
Using the Fourier transform in Eq.~(\ref{scalar}), we have the following action for the photon field,
\begin{eqnarray}
S&=&S_{0} [A]+S_{int} [A,\phi ]\nonumber\\
&=&-\frac{1}{2} \int d\tau \int d^{3} k{\kern 1pt} {\kern 1pt} {\kern 1pt} A^{*} (\vec{k},\tau )\left[\partial _{\tau}^{2} +\vec{k}^{2} - 2aH\xi |\vec{k}| \right] A(\vec{k},\tau )\nonumber\\
&&-\frac{\alpha }{f} \int d\tau \int d^{3} x{\kern 1pt} {\kern 1pt} d^{3} x'd^{3} x''\; \phi (\vec{x}'',\tau )\left(\partial _{\tau} A(\vec{x}',\tau )\right)A(\vec{x},\tau )\mathcal{J} (\vec{x},\vec{x}',\vec{x}''),
\end{eqnarray}
where
\begin{eqnarray}
\mathcal{J} (\vec{x},\vec{x}',\vec{x}'')&=&\int \frac{d^{3}k}{(2\pi )^{3}} \frac{d^{3}k'}{(2\pi )^{3}} {\kern 1pt} \, e^{-i\vec{k}\cdot (\vec{x}-\vec{x}'')} \, e^{-i\vec{k}'\cdot (\vec{x}'-\vec{x}'')}
|\vec{k}|(\vec{\varepsilon }_{R} (\vec{k})\cdot \vec{\varepsilon }_{R} (\vec{k}')) {\rm \; } \nonumber\\
&=&\mathcal{J} (\vec{x}-\vec{x}'',\vec{x}'-\vec{x}'').
\end{eqnarray}
is the interaction kernel. The interaction action can be written in a more compact form as
\begin{equation}
S_{int} [A,\phi ]=\frac{\alpha }{f} \int d^{4} x{\kern 1pt} {\kern 1pt} d^{4} x'd^{4} x''\; A(x)A(x')\phi (x'')\mathcal{F} (x,x',x''),
\label{Sint}
\end{equation}
where
\begin{eqnarray}
\mathcal{F} (x,x',x'')&=&\delta(\tau-\tau'') \partial_{\tau'}\delta(\tau'-\tau'')\mathcal{J} (\vec{x},\vec{x}',\vec{x}'') \nonumber\\
&=&\int \frac{d^{4}k}{(2\pi )^{4}} \frac{d^{4}k'}{(2\pi )^{4}} {\kern 1pt} \, e^{-ik(x-x'')} \, e^{-ik'(x'-x'')}\, i\omega'\,
|\vec{k}|(\vec{\varepsilon }_{R} (\vec{k})\cdot \vec{\varepsilon }_{R} (\vec{k}')) {\rm \; }
\label{mathcalF}
\end{eqnarray}
is the four dimensional interaction kernel.

To implement the CTP formalism we use two copies of both the inflaton and the photon fields, and write the CTP action as
\begin{equation}
S=S_{0} [A_{+} ]-S_{0} [A_{-} ]+S_{int} [A_{+} ,\phi _{+} ]-S_{int} [A_{-} ,\phi _{-} ].
\end{equation}
The corresponding influence action $S_{IF}$ is defined by the path integral over the photon degrees of freedom,
\begin{equation}
e^{iS_{IF} } =\int _{CTP}DA_{+} DA_{-}  e^{iS}.
\label{SIF}
\end{equation}
with the CTP boundary conditions. Since we shall follow the perturbative approach, we add sources to the photon action and express the influence functional $e^{iS_{IF}}$ as
\begin{eqnarray}
&&\int _{CTP}DA_{+} DA_{-}  e^{iS+i\int J_{+} A_{+}  -i\int J_{-} A_{-}  }\nonumber\\
&=&e^{iS_{int} \left[\frac{1}{i} \frac{\delta }{\delta J_{+} } ,\phi _{+} \right]-iS_{int} \left[\frac{1}{-i} \frac{\delta }{\delta J_{-} } ,\phi _{-} \right]} \int _{CTP}DA_{+} DA_{-}  e^{i\left(S_{0} [A_{+} ]-S_{0} [A_{-} ]+\int J_{+} A_{+} -\int J_{-} A_{-}   \right)} %\nonumber
%\\
%&=&\left. e^{iS_{int} \left[\frac{1}{i} \frac{\delta }{\delta J_{+} } ,\phi _{+} \right]-iS_{int} \left[\frac{1}{-i} \frac{\delta }{\delta J_{-} } ,\phi _{-} %\right]} \, e^{-\frac{i}{2} \int \left(J_{+} G_{++} J_{+} -J_{+} G_{+-} J_{-} -J_{-} G_{-+} J_{+} +J_{-} G_{--} J_{-} \right) } \right|_{J_{+} =J_{-} =0}.
\label{expSIF}
\end{eqnarray}
This CTP integral can be evaluated using the Schwinger-Keldysh Green's functions of the photon field $A(\vec{x},\tau)$. As a scalar field $A(\vec{x},\tau )$ can be written in canonical form,
\begin{equation}
\hat{A}(\vec{x},\tau )=\int \frac{d^{3} k}{(2\pi )^{3/2} }  {\kern 1pt} \,
\left[\hat{a}(\vec{k}) A (\vec{k},\tau ) e^{i\vec{k}\cdot \vec{x}} + {\rm h.c.}\right],
\end{equation}
where the mode function $A (\vec{k},\tau )$ satisfies
\begin{equation}
\left[\partial _{\tau}^{2} +\vec{k}^{2} - 2aH\xi |\vec{k}| \right]A(\vec{k},\tau )=0.
\end{equation}
It is chosen in such a way that as $\tau\rightarrow -\infty$ it coincides with the Minkowski vacuum mode function (see Appendix \ref{mode}). The explicit form of this mode function will be given in the next section. Then the Schwinger-Keldysh Green's functions can be expressed in terms of $A (\vec{k},\tau )$.
In particular, the Green's function $G_{++} (x,x')$ can be written as
\begin{eqnarray}
&&-iG_{++} (x,x')\nonumber\\
%&=&\left.\frac{\delta }{\delta J_{+}(x) }\frac{\delta }{\delta J_{+}(x') } e^{-\frac{i}{2} \int \left(J_{+} G_{++} J_{+} -J_{+} G_{+-} J_{-} -J_{-} G_{-+} J_{+} +J_{-} G_{--} J_{-} \right) } \right|_{J_{+} =J_{-} =0}\nonumber\\
&=&\left.\frac{\delta }{\delta J_{+}(x) }\frac{\delta }{\delta J_{+}(x') } \int _{CTP}DA_{+} DA_{-}  e^{i\left(S_{0} [A_{+} ]-S_{0} [A_{-} ]+\int J_{+} A_{+} -\int J_{-} A_{-} \right)} \right|_{J_{+} =J_{-} =0}\nonumber\\
&=& -\langle T\hat{A}_{+}(x)\hat{A}_{+}(x')\rangle.
\end{eqnarray}
Hence, in terms of the mode function $A(\vec{k},\tau)$, this Green's function can be expressed as
\begin{eqnarray}
G_{++} (x,x')&=&-i\theta (\tau -\tau ')\int \frac{d^{3} k}{(2\pi )^{3} }  {\kern 1pt} \, A(\vec{k},\tau )A^{*} (\vec{k},\tau '){\kern 1pt} {\kern 1pt} e^{i\vec{k}\cdot (\vec{x}-\vec{x}')} \nonumber\\
&&-i\theta (\tau '-\tau )\int \frac{d^{3} k}{(2\pi )^{3} }  {\kern 1pt} \, A(\vec{k},\tau ')A^{*} (\vec{k},\tau ){\kern 1pt} {\kern 1pt} e^{-i\vec{k}\cdot (\vec{x}-\vec{x}')}.\label{greenpp}
\end{eqnarray}
Similarly for the other Green's functions, we have
\begin{eqnarray}
G_{+-} (x,x')&=&-i\left.\frac{\delta }{\delta J_{+}(x) }\frac{\delta }{\delta J_{-}(x') } \int _{CTP}DA_{+} DA_{-}  e^{i\left(S_{0} [A_{+} ]-S_{0} [A_{-} ]+\int J_{+} A_{+} -\int J_{-} A_{-} \right)} \right|_{J_{+} =J_{-} =0}\nonumber\\
&=&-i\langle \hat{A}_{-}(x')\hat{A}_{+}(x)\rangle\nonumber\\
&=&-i\int \frac{d^{3} k}{(2\pi )^{3} }  {\kern 1pt} \, A(\vec{k},\tau ')A^{*} (\vec{k},\tau ){\kern 1pt} {\kern 1pt} e^{-i\vec{k}\cdot (\vec{x}-\vec{x}')},\\
G_{-+} (x,x')&=&-i\left.\frac{\delta }{\delta J_{+}(x') }\frac{\delta }{\delta J_{-}(x) } \int _{CTP}DA_{+} DA_{-}  e^{i\left(S_{0} [A_{+} ]-S_{0} [A_{-} ]+\int J_{+} A_{+} -\int J_{-} A_{-} \right)} \right|_{J_{+} =J_{-} =0}\nonumber\\
&=&-i\langle \hat{A}_{-}(x)\hat{A}_{+}(x')\rangle\nonumber\\
&=&-i\int \frac{d^{3} k}{(2\pi )^{3} }  {\kern 1pt} \, A(\vec{k},\tau )A^{*} (\vec{k},\tau '){\kern 1pt} {\kern 1pt} e^{i\vec{k}\cdot (\vec{x}-\vec{x}')}\nonumber\\
&=&G_{+-} (x',x), \\
G_{--} (x,x')&=&i\left.\frac{\delta }{\delta J_{-}(x) }\frac{\delta }{\delta J_{-}(x') } \int _{CTP}DA_{+} DA_{-}  e^{i\left(S_{0} [A_{+} ]-S_{0} [A_{-} ]+\int J_{+} A_{+} -\int J_{-} A_{-} \right)} \right|_{J_{+} =J_{-} =0}\nonumber\\
&=&-i\langle \bar{T}\hat{A}_{-}(x)\hat{A}_{-}(x')\rangle\nonumber\\
&=&-i\theta (\tau -\tau ')\int \frac{d^{3} k}{(2\pi )^{3} }  {\kern 1pt} \, A(\vec{k},\tau ')A^{*} (\vec{k},\tau ){\kern 1pt} {\kern 1pt} e^{-i\vec{k}\cdot (\vec{x}-\vec{x}')} \nonumber\\
&&-i\theta (\tau '-\tau )\int \frac{d^{3} k}{(2\pi )^{3} }  {\kern 1pt} \, A(\vec{k},\tau )A^{*} (\vec{k},\tau '){\kern 1pt} {\kern 1pt} e^{i\vec{k}\cdot (\vec{x}-\vec{x}')} \nonumber\\
&=&-G_{++}^{*} (x,x')\label{greenmm}
\end{eqnarray}
where $\bar{T}$ is for the anti-time ordered operation.

>From the representations of these Green's functions in terms of the functional derivatives with respect to the sources, it is possible to express the zeroth order CTP integral in Eq.~(\ref{expSIF}) as
\begin{eqnarray}
&&\int _{CTP}DA_{+} DA_{-}  e^{i\left(S_{0} [A_{+} ]-S_{0} [A_{-} ]+\int J_{+} A_{+} -\int J_{-} A_{-}   \right)}\nonumber\\
&=&e^{-\frac{i}{2} \int \left(J_{+} G_{++} J_{+} -J_{+} G_{+-} J_{-} -J_{-} G_{-+} J_{+} +J_{-} G_{--} J_{-} \right) }\label{0orderCTP}
\end{eqnarray}
Putting this result back into Eqs.~(\ref{SIF}) and (\ref{expSIF}) the influence functional $S_{IF}$ can be evaluated in a perturbative manner in powers of $\alpha/f$, the parameter in the interaction action. The first-order term gives the in-in expectation value of the interaction. Up to the second order in the inflaton field it is possible to identify the corresponding noise and dissipation kernels due to the inflaton-photon interaction. We shall work this out in detail in the next two sections. Subsequently one can also obtain the equation of motion of the inflaton field by taking functional variation on the influence action. This equation of motion will be in the form of a Langevin equation with a stochastic force of which the correlation function is given by the noise kernel.

\section{Influence functional}

In this section we shall detail the evaluation of the influence action $S_{IF}$ in powers of the parameter $\alpha/f$. As shown in the last section this is achieved by taking functional derivatives on the zeroth order CTP integral which is given in terms of the various Schwinger-Keldysh Green's functions. These functions are constructed using the mode function of the corresponding quantum field (Eqs. (\ref{greenpp}) to (\ref{greenmm})). In the case where $\xi$ is large as compared to $|k\tau|$, the right-handed photon $A_{R}(k,\tau)$ is enhanced while the left-handed one is not. Therefore we can use the large $\xi$ approximation of $A_{R}(k,\tau)$ (Eq.~(\ref{modesol})) for mode function $A(\vec{k},\tau)$. Dropping the constant angular phase in $A_{R}(k,\tau)$, we have
\begin{equation}
A(k,\tau) = \frac{1}{\sqrt{2k}}\left(\frac{-k\tau}{2\xi}\right)^{1/4} e^{\pi\xi} e^{-2\sqrt{-2\xi k\tau}}.
\label{Amodesol}
\end{equation}
Note that $A(\vec{k},\tau)$ is real and depends only on $k$. As a consequence, all the Schwinger-Keldysh Green's functions are equal:
\begin{eqnarray}
&&G_{++} (x,x')=G_{--} (x,x')=G_{+-} (x,x')=G_{-+} (x',x)=G_{-+} (x,x')\equiv G(x,x')\,;\nonumber\\
&&G(x,x')=-i\int \frac{d^{3} k}{(2\pi )^{3} }  {\kern 1pt} \, A(k,\tau) A(k,\tau'){\kern 1pt} {\kern 1pt} e^{i\vec{k}\cdot (\vec{x}-\vec{x}')}.
\label{greenfct}
\end{eqnarray}

%In the subsequent calculation we shall encounter integrals involving $G(x,x')$. Since we are only interested in the values of these integrals in the large $\xi$ limit, it is convenient to have a general formula for the asymptotic expansion in $1/\xi$ for the following integral.
%\begin{eqnarray}
%&&\int\frac{d^{3}k}{(2\pi)^{3}}k^{\alpha}e^{i\vec{k}\cdot\Delta\vec{x}}A(\vec{k},\tau)A(\vec{k},\tau')\nonumber\\
%&=&\frac{(\tau\tau')^{1/4}}{\pi^{2}}e^{2\pi\xi}\sum_{n=0}^{\infty}\frac{(-1)^{n}}{\xi^{\alpha+2n+3}}
%\left(\frac{\Gamma(2\alpha+4n+5)}{2^{3\alpha+6n+9}\Gamma(2n+2)}\right)\frac{(\Delta x)^{2n}}{(\sqrt{-\tau}+\sqrt{-\tau'})^{2\alpha+4n+5}}\label{asyexp}
%\end{eqnarray}
%where $\Delta x=|\Delta\vec{x}|$. Note that this asymptotic expansion is also a small $\Delta x$ expansion.

Now we are ready to expand the influence functional $e^{iS_{IF}}$~(\ref{expSIF}) in powers of $\alpha/f$ in $S_{int}$~(\ref{Sint}),
\begin{eqnarray}
S_{IF}=\sum_{i=1}^{\infty}\delta S^{(i)}_{IF}.\label{SIFexp}
\end{eqnarray}
Following Eq.~(\ref{expSIF}) the first order $\delta S^{(1)}_{IF}$ is obtained by second functional derivatives on the zeroth order CTP integral (Eq.~(\ref{0orderCTP})).
\begin{eqnarray}
\delta S^{(1)}_{IF}&=&\frac{\alpha }{f}\int d^{4} x{\kern 1pt} {\kern 1pt} d^{4} x'd^{4} x''\; \left[\frac{1}{i} \frac{\delta }{\delta J_{+}(x)} \frac{1}{i} \frac{\delta }{\delta J_{+} (x')} \phi_{+} (x'')- \frac{1}{-i} \frac{\delta }{\delta J_{-}(x)} \frac{1}{-i} \frac{\delta }{\delta J_{-} (x')} \phi_{-} (x'')\right] \nonumber \\
&& \left.\times\; \mathcal{F} (x,x',x'')
e^{-\frac{i}{2} \int \left(J_{+} G_{++} J_{+} -J_{+} G_{+-} J_{-} -J_{-} G_{-+} J_{+} +J_{-} G_{--} J_{-} \right) } \right|_{J_{+} =J_{-} =0}  \nonumber \\
%&=& -\frac{\alpha }{f} \int d^{4} x{\kern 1pt} {\kern 1pt} d^{4} x'd^{4} x''\; \mathcal{F} (x,x',x'') \left[G_{++}(x,x') \phi_{+} (x'') -G_{--}(x,x') \phi_{-} %(x'') \right] \nonumber\\
&=& i\frac{\alpha }{f} \int d^{4} x{\kern 1pt} {\kern 1pt} d^{4} x'd^{4} x''\; \mathcal{F} (x,x',x'') G(x,x') \Delta\phi(x''),
\end{eqnarray}
where we have defined $\Delta\phi(x)\equiv \phi_{+}(x)-\phi_{-}(x)$.
Using Eqs.~(\ref{polvec}), (\ref{mathcalF}), and (\ref{greenfct}), it is straightforward to show that
\begin{equation}
\int d^{4} x{\kern 1pt} {\kern 1pt} d^{4} x'\; \mathcal{F} (x,x',x'') G(x,x')
={i\over 2} \partial_{\tau''}\int \frac{d^3 k}{(2\pi)^3} \,k\, A(k,\tau'')  A(k,\tau'').
\end{equation}
where $A(k,\tau)$ is given by the large $\xi$ approximation in Eq.~(\ref{Amodesol}). 
In the leading order of the asymptotic expansion of $1/\xi$, this integral can be evaluated readily to give
\begin{eqnarray}
\int d^{4} x{\kern 1pt} {\kern 1pt} d^{4} x'\; \mathcal{F} (x,x',x'') G(x,x')
=i\left(\frac{135}{65536\pi^2}\right)\frac{e^{2\pi\xi}}{\xi^{4}\tau^{4}}
\end{eqnarray}
%Hence, using the result in Eq.~(\ref{asyexp}) for $\alpha=1$ and $\Delta x=0$, 
Hence, we reach the final form of the first-order term as
\begin{equation}
\delta S_{IF}^{(1)}=-\left(\frac{\alpha}{f}e^{2\pi\xi}\right) \int d^{4} x{\kern 1pt} {\kern 1pt}\Delta\phi(x) \frac{135}{65536\pi^2\xi^4\tau^4}.
\end{equation}

For the second-order term, we have the following functional derivatives.
\begin{eqnarray}
&&\frac{\alpha^2}{2f^2}\left\{i\int d^{4} x{\kern 1pt} {\kern 1pt} d^{4} x'd^{4} x''\; \left[\frac{1}{i} \frac{\delta }{\delta J_{+}(x)} \frac{1}{i} \frac{\delta }{\delta J_{+} (x')} \phi_{+} (x'')- \frac{1}{-i} \frac{\delta }{\delta J_{-}(x)} \frac{1}{-i} \frac{\delta }{\delta J_{-} (x')} \phi_{-} (x'')\right] \right.\nonumber \\
&& \left.\left. \mathcal{F} (x,x',x'') \right\}^2
e^{-\frac{i}{2} \int \left(J_{+} G_{++} J_{+} -J_{+} G_{+-} J_{-} -J_{-} G_{-+} J_{+} +J_{-} G_{--} J_{-} \right) } \right|_{J_{+} =J_{-} =0}  \nonumber \\
&=&\frac{\alpha^{2}}{2f^{2}}\int\Delta\phi(x_{5})\left\{\mathcal{F}(x_{1},x_{2},x_{5})
\left[G(x_{1},x_{2})G(x_{3},x_{4})+G(x_{1},x_{3})G(x_{2},x_{4})\right.\right.\nonumber\\
&&\hskip 80pt \ \ \ \ \ \ \ \ \ \ \ \ \ \ \ \ \ \ \ \  \left.\left.+G(x_{1},x_{4})G(x_{2},x_{3})\right]\mathcal{F}(x_{3},x_{4},x_{6})\right\}\Delta\phi(x_{6}).\label{secord}
\end{eqnarray}
The first term in the brackets above represents the disconnected diagram which we shall ignore. For the other two terms we follow the asymptotic analysis in~\cite{benderorszag} to obtain the leading contribution in $1/\xi$ in the evaluation of the corresponding integrals. We then again arrived at the following asymptotic expansions:
\begin{eqnarray}
&&\int d^{4}x_{1}d^{4}x_{2}d^{4}x_{3}d^{4}x_{4}\mathcal{F}(x_{1},x_{2},x_{5})G(x_{1},x_{3})G(x_{2},x_{4})\mathcal{F}(x_{3},x_{4},x_{6})\nonumber\\
&=&\int\frac{d^{3}k}{(2\pi)^{3}}e^{i\vec{k}\cdot(\vec{x}_{5}-\vec{x}_{6})}e^{4\pi\xi}\left[-\left(\frac{15\,k^{3/2}}{256\sqrt{2}\,\pi^{2}\,\xi^{7/2}}\right)\frac{1}{(\sqrt{-\tau_{5}}+\sqrt{-\tau_{6}})^{7}}\,e^{-2\sqrt{2\xi k}(\sqrt{-\tau_{5}}+\sqrt{-\tau_{6}})}
+\dots\right],\nonumber\\
%\left[1-\frac{495}{16(\sqrt{-\tau_{5}}+\sqrt{-\tau_{6}})^{4}\xi^{2}}|\vec{x}_{5}-\vec{x}_{6}|^{2}+\cdots\right],\nonumber\\
\\
&&\int d^{4}x_{1}d^{4}x_{2}d^{4}x_{3}d^{4}x_{4}\mathcal{F}(x_{1},x_{2},x_{5})G(x_{1},x_{3})G(x_{2},x_{4})\mathcal{F}(x_{4},x_{3},x_{6})\nonumber\\
&=&\int\frac{d^{3}k}{(2\pi)^{3}}e^{i\vec{k}\cdot(\vec{x}_{5}-\vec{x}_{6})}e^{4\pi\xi}\left[-\left(\frac{105\,k}{1024\,\pi^{2}\,\xi^{4}}\right) \frac{1}{(\sqrt{-\tau_{5}}+\sqrt{-\tau_{6}})^{8}} \,e^{-2\sqrt{2\xi k}(\sqrt{-\tau_{5}}+\sqrt{-\tau_{6}})}+\dots\right].
%\left[1-\frac{111}{4(\sqrt{-\tau_{5}}+\sqrt{-\tau_{6}})^{4}\xi^{2}}|\vec{x}_{5}-\vec{x}_{6}|^{2}+\cdots\right].\nonumber\\
\end{eqnarray}
With these expansions the second order term in Eq.~(\ref{SIFexp}) can be expressed as
\begin{eqnarray}
\delta S_{IF}^{(2)}&=&i\left(\frac{\alpha^{2}}{2f^{2}}e^{4\pi\xi}\right)\int d^{4}x\, d^{4}x'\Delta\phi(x)\Delta\phi(x')\,
\int\frac{d^{3}k}{(2\pi)^{3}}e^{i\vec{k}\cdot(\vec{x}-\vec{x}')}e^{-2\sqrt{2\xi k}(\sqrt{-\tau}+\sqrt{-\tau'})}\nonumber\\
&&\Bigg\{\left[\left(\frac{15\,k^{3/2}}{256\sqrt{2}\,\pi^{2}\,\xi^{7/2}}\right)\frac{1}{(\sqrt{-\tau}+\sqrt{-\tau'})^{7}}+\left(\frac{105\,k}{1024\,\pi^{2}\,\xi^{4}}\right) \frac{1}{(\sqrt{-\tau}+\sqrt{-\tau'})^{8}}\right]+\dots\Bigg\}\nonumber\\
%&&\hskip 100pt\left.\left[1-\frac{589}{20(\sqrt{-\tau}+\sqrt{-\tau'})^{4}\xi^{2}}|\vec{x}-\vec{x}'|^{2}+\cdots\right]\right\}\Delta\phi(x').
\end{eqnarray}
where the ellipsis represents terms higher order in $1/\xi$ in the asymptotic expansion.

Similarly, omitting the disconnected terms, we can write the third-order term as
\begin{eqnarray}
&&\frac{\alpha^3}{6f^3}\left\{i\int d^{4} x{\kern 1pt} {\kern 1pt} d^{4} x'd^{4} x''\; \left[\frac{1}{i} \frac{\delta }{\delta J_{+}(x)} \frac{1}{i} \frac{\delta }{\delta J_{+} (x')} \phi_{+} (x'')- \frac{1}{-i} \frac{\delta }{\delta J_{-}(x)} \frac{1}{-i} \frac{\delta }{\delta J_{-} (x')} \phi_{-} (x'')\right] \right.\nonumber \\
&& \left.\left. \mathcal{F} (x,x',x'') \right\}^3
e^{-\frac{i}{2} \int \left(J_{+} G_{++} J_{+} -J_{+} G_{+-} J_{-} -J_{-} G_{-+} J_{+} +J_{-} G_{--} J_{-} \right) } \right|_{J_{+} =J_{-} =0}  \nonumber \\
&=&-\frac{\alpha^{3}}{6f^{3}} \int\Delta\phi(x_{7})\Delta\phi(x_{8})\Delta\phi(x_{9})
\mathcal{F}(x_{1},x_{2},x_{7})\mathcal{F}(x_{3},x_{4},x_{8})\mathcal{F}(x_{5},x_{6},x_{9}) \nonumber \\
&&\ \ \ \ \ \ \ \ \ \ \ \ \left[G(x_{1},x_{3})G(x_{2},x_{5})G(x_{4},x_{6})+G(x_{1},x_{3})G(x_{2},x_{6})G(x_{4},x_{5})\right. \nonumber \\
&&\ \ \ \ \ \ \ \ \ \ \ \ +G(x_{1},x_{4})G(x_{2},x_{6})G(x_{3},x_{5})+G(x_{1},x_{5})G(x_{2},x_{3})G(x_{4},x_{6}) \nonumber \\
&&\ \ \ \ \ \ \ \ \ \ \ \ +G(x_{1},x_{5})G(x_{2},x_{4})G(x_{3},x_{6})+G(x_{1},x_{6})G(x_{2},x_{4})G(x_{3},x_{5})\nonumber\\
&&\ \ \ \ \ \ \ \ \ \ \ \ \left.+G(x_{1},x_{4})G(x_{2},x_{5})G(x_{3},x_{6})+G(x_{1},x_{6})G(x_{2},x_{3})G(x_{4},x_{5}) \right].\label{3order}
\label{third}
\end{eqnarray}
>From the symmetry of the terms we can separate them into two groups, one being with the first six terms and the other with the last two terms. Note that in the first group, two of the coordinates, $x_{2}$, $x_{4}$, and $x_{6}$, are in one Green's function. In the last two terms, these coordinates are present in the three different Green's functions. 

Let us evaluate the integral involving the first term. Following similar steps in dealing with the terms in $\delta S_{IF}^{(2)}$, we obtain
\begin{eqnarray}
&&\int d^4x_1  d^4x_2  d^4x_3  d^4x_4  d^4x_5  d^4x_6 \mathcal{F}(x_{1},x_{2},x_{7})\mathcal{F}(x_{3},x_{4},x_{8})\mathcal{F}(x_{5},x_{6},x_{9})\nonumber\\
&&\ \ \ \ \ G(x_{1},x_{3})G(x_{2},x_{5})G(x_{4},x_{6}) \nonumber \\
&=&\int\frac{d^{3}k}{(2\pi)^{3}}\int\frac{d^{3}k'}{(2\pi)^{3}}\,e^{i\vec{k}\cdot(\vec{x}_{7}-\vec{x}_{8})}\,e^{i\vec{k}'\cdot(\vec{x}_{8}-\vec{x}_{9})}e^{6\pi\xi}\nonumber\\
&&\Bigg[(-i)\left( \frac{105\,k^{3/2}k'^{1/2}}{8192\pi^2\xi^{4}}\right)\left(1+\frac{\vec{k}\cdot\vec{k}'}{kk'}\right)^{2}\
\frac{e^{-2\sqrt{2\xi}[(\sqrt{-\tau_{7}}+\sqrt{-\tau_{8}})\sqrt{k}+(\sqrt{-\tau_{8}}+\sqrt{-\tau_{9}})\sqrt{k'}]}}{(\sqrt{-\tau_7}+\sqrt{-\tau_9})^{8}}+\dots\Bigg]\nonumber\\
%&&\ \ \ \ \ \left\{ 1+\frac{1}{\xi^2}\left[\frac{135}{8}\left(\frac{|\vec{x}_7-\vec{x}_8||\vec{x}_7-\vec{x}_9|}{(\sqrt{-\tau_7}
%+\sqrt{-\tau_8})^2(\sqrt{-\tau_7}+\sqrt{-\tau_9})^2}\right)\right.\right.\nonumber\\
%&&\ \ \ \ \ \ \ \ -\frac{105}{8}\left(\frac{|\vec{x}_7-\vec{x}_8||\vec{x}_8-\vec{x}_9|}{(\sqrt{-\tau_7}+\sqrt{-\tau_8})^2(\sqrt{-\tau_8}+\sqrt{-\tau_9})^2}\right)
%\right.
% \nonumber \\
%&&\ \ \ \ \ \ \ \ \left.\left.
%+\frac{21}{2}\left(\frac{|\vec{x}_7-\vec{x}_9||\vec{x}_8-\vec{x}_9|}{(\sqrt{-\tau_7}+\sqrt{-\tau_9})^2(\sqrt{-\tau_8}+\sqrt{-\tau_9})^2}\right)\right] +\cdots\right\}.
\end{eqnarray}
The other five integrals in the same group can be easily worked out by making permutations of $\vec{x}_7$, $\vec{x}_8$, and $\vec{x}_9$. The last two integrals in Eq.~(\ref{3order}) can be similarly evaluated. Hence, the third-order term in Eq.~(\ref{SIFexp}) is
\begin{eqnarray}
&&\delta S_{IF}^{(3)}\nonumber\\
&=&\left(\frac{\alpha^{3}}{6f^{3}}e^{6\pi\xi}\right)\int d^4x_{1}\,  d^4x'_{2}\,  d^4x_{3}\, \Delta\phi(x_{1})\Delta\phi(x_{2})\Delta\phi(x_{3})\nonumber\\
&&\int\frac{d^{3}k}{(2\pi)^{3}}\int\frac{d^{3}k'}{(2\pi)^{3}}\,e^{i\vec{k}\cdot(\vec{x}_{1}-\vec{x}_{2})}\,e^{i\vec{k}'\cdot(\vec{x}_{2}-\vec{x}_{3})}
\times\nonumber\\
&&\Bigg\{\left[ \frac{105\,(k^{3/2}k'^{1/2}+kk'/3)}{8192\pi^2\xi^{4}}\right]\left(1+\frac{\vec{k}\cdot\vec{k}'}{kk'}\right)^{2}\
\frac{e^{-2\sqrt{2\xi}[(\sqrt{-\tau_{1}}+\sqrt{-\tau_{2}})\sqrt{k}+(\sqrt{-\tau_{2}}+\sqrt{-\tau_{3}})\sqrt{k'}]}}{(\sqrt{-\tau_1}+\sqrt{-\tau_3})^{8}}\nonumber\\
&&\ \ +\dots\Bigg\}
%&&+\left. \frac{3}{4\xi^2}\left[T^{(1)}(\sqrt{-\tau_7},\sqrt{-\tau_8},\sqrt{-\tau_9}) |\vec{x}_7-\vec{x}_9||\vec{x}_8-\vec{x}_9| + {\rm perm} (x_7,x_8,x_9)\right]+\cdots\right\},
\end{eqnarray}
%where the function
%\begin{eqnarray}
%&&T(a_1,a_2,a_3) \nonumber \\
%&=& \frac{8(a_1^3+a_2^3+a_3^3)+27(a_1^2a_2+a_1a_2^2+a_2^2a_3+a_2a_3^2+a_3^2a_1+a_1^2a_3)+62a_1a_2a_3}
%{(a_1+a_2)^{9}(a_2+a_3)^{9}(a_3+a_1)^{9}} \,.\label{T0}
%&&T^{(1)}(a_1,a_2,a_3) \nonumber \\
%&=& \frac{10(a_1^3+a_2^3)+56a_3^2+149(a_1a_2^2+a_2a_1^2)+69(a_1^2+a_2^2)a_3+35(a_1+a_2)a_3^2+194a_1a_2a_3}
%{(a_1+a_2)^{9}(a_2+a_3)^{11}(a_3+a_1)^{11}}  \,.\nonumber\\ \label{T1}
%\end{eqnarray}

%\section{Noise kernel}

Let us summarize what we have obtained for the perturbative expansion of the influence action $S_{IF}$. Up to the third order, we have
\begin{eqnarray}
&&S_{IF}\nonumber\\
&=& -\int d^4x \sqrt{-g(x)} \delta V(x) \Delta\phi(x) \nonumber\\
&&\ \ +{i\over 2} \int d^4x_1 d^4x_2 \sqrt{-g(x_1)} \sqrt{-g(x_2)} \Delta\phi(x_1) N_2(x_1,x_2) \Delta\phi(x_2)
\nonumber \\
&&\ \ +{1\over 6} \int d^4x_1 d^4x_2 d^4x_3 \sqrt{-g(x_1)} \sqrt{-g(x_2)} \sqrt{-g(x_3)}\Delta\phi(x_1)\Delta\phi(x_2)\Delta\phi(x_3) N_3(x_1,x_2,x_3)
\nonumber\\ &&\ \ +\cdots
\label{SIFfinal}
\end{eqnarray}
where
\begin{eqnarray}
\delta V(x)
 &=& \frac{1}{a(\tau)} \left(\frac{\alpha}{f}e^{2\pi\xi}\right) \left(\frac{135}{65536\pi^2}\right) \frac{1}{\xi^4\tau^4}
\end{eqnarray}
\begin{eqnarray}
&&N_2(x_1,x_2) \nonumber\\
&=&  \frac{1}{a^4(\tau_1)a^4(\tau_2)} \left(\frac{\alpha^2}{f^2}e^{4\pi\xi}\right)
\int\frac{d^{3}k}{(2\pi)^{3}}e^{i\vec{k}\cdot(\vec{x}_{1}-\vec{x}_{2})}e^{-2\sqrt{2\xi k}(\sqrt{-\tau_{1}}+\sqrt{-\tau_{2}})}\nonumber\\
&&\Bigg\{\left[\left(\frac{15\,k^{3/2}}{256\sqrt{2}\,\pi^{2}\,\xi^{7/2}}\right)\frac{1}{(\sqrt{-\tau_{1}}+\sqrt{-\tau_{2}})^{7}}+\left(\frac{105\,k}{1024\,\pi^{2}\,\xi^{4}}\right) \frac{1}{(\sqrt{-\tau_{1}}+\sqrt{-\tau_{2}})^{8}}\right]+\dots\Bigg\}\nonumber\\
\label{noisekernel}
\end{eqnarray}
%&&\left[1-\frac{589}{20(\sqrt{-\tau_1}+\sqrt{-\tau_2})^{4}\xi^{2}}|\vec{x}_1-\vec{x}_2|^{2}+\cdots\right]\,, \label{noisekernel} \\
\begin{eqnarray}
&&N_3(x_1,x_2,x_3)\nonumber\\
&=& \frac{1}{a^4(\tau_1)a^4(\tau_2)a^4(\tau_3)}
\left(\frac{\alpha^{3}}{f^{3}}e^{6\pi\xi}\right) \times\nonumber\\
&&\int\frac{d^{3}k}{(2\pi)^{3}}\int\frac{d^{3}k'}{(2\pi)^{3}}\,e^{i\vec{k}\cdot(\vec{x}_{1}-\vec{x}_{2})}\,e^{i\vec{k}'\cdot(\vec{x}_{2}-\vec{x}_{3})}
\times\nonumber\\
&&\Bigg\{\left[ \frac{105\,(k^{3/2}k'^{1/2}+kk'/3)}{8192\pi^2\xi^{4}}\right]\left(1+\frac{\vec{k}\cdot\vec{k}'}{kk'}\right)^{2}\
\frac{e^{-2\sqrt{2\xi}[(\sqrt{-\tau_{1}}+\sqrt{-\tau_{2}})\sqrt{k}+(\sqrt{-\tau_{2}}+\sqrt{-\tau_{3}})\sqrt{k'}]}}{(\sqrt{-\tau_1}+\sqrt{-\tau_3})^{8}}\nonumber\\
&&\ \ +\,5\ {\rm permutations\ of\ }(x_{1},x_{2},x_{3})+ \dots\Bigg\}
%&&+ \frac{3}{4\xi^2}\left[T^{(1)}(\sqrt{-\tau_1},\sqrt{-\tau_2},\sqrt{-\tau_3}) |\vec{x}_1-\vec{x}_3||\vec{x}_2-\vec{x}_3|\right.\nonumber\\
%&&\left.\left.+ {\rm perm} (x_1,x_2,x_3)\right]+\cdots\right\},
\label{Nthree}
\end{eqnarray}

%\section{Noise kernel}
\section{Noise kernel and the Langevin equation}

The influence action $S_{IF}$ obtained in the previous section has a stochastic interpretation in analogue to the quantum Brownian motion model~\cite{calzettahu}. In particular, the terms quadratic and cubic in $\Delta\phi$ in $S_{IF}$ can be represented as the effect of a stochastic force.

In the usual approach, where the cubic term is absent, the influence action can be rewritten as a path integral over a stochastic force $\zeta$.
\begin{equation}
e^{-\frac{1}{2}\int \Delta\phi\,N_2\,\Delta\phi}\equiv\int D\zeta\,P_{g}[\zeta]\,e^{-i\int\,\zeta\,\Delta\phi},
\end{equation}
where $P_{g}[\zeta]$ is a gaussian probability functional
\begin{equation}
P_{g}[\zeta]={\rm det}(2\pi N_{2})^{-1/2}\,e^{-\frac{1}{2}\int\,\zeta\,N_{2}^{-1}\,\zeta}.
\end{equation}
With this probability functional one obtain the correlation function of the stochastic force $\zeta$
\begin{equation}
\langle \zeta(x_1)\zeta(x_2) \rangle= \int D\zeta \,P_{g}[\zeta]\,\zeta(x_1)\,\zeta(x_2) =N_2(x_1,x_2).
\end{equation}
Therefore, $N_{2}(x_1,x_2)$ is referred to as the noise kernel characterizing the quantum fluctuations of the environment, that is, the photon field.

Now with the quadratic and the cubic $\Delta\phi$ terms in Eq.~(\ref{SIFfinal}), it is still possible to rewrite them as the effect of a single stochastic force.
\begin{equation}
e^{-{1\over2}\int \Delta\phi\, N_2\, \Delta\phi+{i\over6}\int\Delta\phi \,\Delta\phi\, \Delta\phi\, N_3} = \int D\zeta \,P[\zeta]\, e^{-i\int \zeta\, \Delta\phi},\label{stochasticpathintegral}
\end{equation}
where $P[\zeta]$ is again a probability distribution function normalized to $\langle 1 \rangle_\zeta \equiv \int D\zeta P[\zeta] =1$. Evaluating the various correlation functions of $\zeta$, we have
\begin{eqnarray}
\langle \zeta(x) \rangle &=&\int D\zeta P[\zeta] \zeta(x) =0,  \\
\langle \zeta(x_1)\zeta(x_2) \rangle&=& \int D\zeta P[\zeta] \zeta(x_1)\zeta(x_2) =N_2(x_1,x_2), \label{N2}\\
\langle \zeta(x_1)\zeta(x_2) \zeta(x_3) \rangle&=& \int D\zeta P[\zeta] \zeta(x_1)\zeta(x_2)\zeta(x_3)=N_3(x_1,x_2,x_3).\label{N3}
\end{eqnarray}
The outstanding feature of these correlators is that the three-point correlation function is nonvanishing. The probability functional $P[\zeta]$ of the stochastic force is therefore nongaussian. As we shall see in the next section, this will be the source of nongaussianity for the fluctuations of the inflaton field.

As in the usual case one can identify $N_2(x_{1},x_{2})$ as the noise kernel originated from the electromagnetic interaction. According to the dissipation-fluctuation theorem, one would expect a corresponding dissipation kernel. However, it is absent from the influence action in Eq.~(\ref{SIFfinal}). In fact, the dissipation and noise kernels are derived from the real and the imaginary parts of the Green's function of the quantum field, respectively. In the large $\xi$ limit, the mode function in Eq.~(\ref{Amodesol}) is actually real, and the corresponding Green's function in Eq.~(\ref{greenfct}) is therefore purely imaginary. As given in Eq.~(\ref{noisekernel}), the noise kernel in  $N_{2}(x_{1},x_{2})\sim e^{4\pi\xi}$. This indicates that the instability of the photon enhances its quantum fluctuations, while the dissipation effects are not enhanced in comparison to that. This could be a general phenomenon which is worth further investigation.

%\section{Langevin equation}

Next, we investigate the corresponding equation of motion for the inflaton field with the effects of the noise kernel included. From the action for the inflaton field $\phi$ in Eq.~(\ref{Stot}) up to the quadratic order,
\begin{equation}
S_\phi = \int d^4x \sqrt{-g} \left[ -{1\over2} \left(\partial \phi \right)^{2} - {1\over2} V''(\Phi ) \phi^2\right].
\end{equation}
Hence, doing the variation
\begin{equation}
\frac{\delta}{\delta\phi_+} \left( S_\phi[\phi_+]-S_\phi[\phi_-]+S_{IF}[\phi_+,\phi_-]\right) {\Big |}_{\phi_+=\phi_-=\phi} = 0,
\end{equation}
the equation of motion for $\phi$ can be obtained as, with $S_{IF}$ given by Eqs.~(\ref{SIF}) and (\ref{stochasticpathintegral}),
\begin{equation}
\frac{1}{\sqrt{-g}} \partial_\mu \left( \sqrt{-g} g^{\mu\nu} \partial_\nu \phi \right) - V''_{ren}(\Phi ) \phi = \zeta,
\end{equation}
which is in the form of a Klein-Gordon-Langevin equation with the stochastic force term given by $\zeta$. Note that we have defined $V''_{ren}(\Phi)=V''(\Phi)+\delta V$. In the metric~(\ref{flatmetric}), it reads
\begin{equation}
\partial_\tau^2 \phi + 2 H \partial_\tau \phi - \nabla^2\phi + a^2 V''_{ren}(\Phi ) \phi = -a^2\zeta.
\end{equation}
We can see that the effect of the photon field, other than the renormalization of the potential, is the production of the stochastic force $\zeta$. The mode function of this equation without the stochastic force term can be solved in the usual manner by choosing the Bunch-Davies vacuum. One can construct the retarded Green's function from this mode function. With this Green's function, the mode function with the stochastic force in the inhomogeneous equation can be obtained. Then we can derive the correlation functions of the inflaton field by taking the expectation value of the inflaton field as well as the stochastic average. However, one can directly work with the path integral by putting in source terms. We shall derive the correlation functions in this approach in the next section.

\section{Correlation functions}

Now we work on the correlation functions of the perturbation $\phi$. To do that we go back to the path integral. We add source terms so that the correlation functions can be obtained by taking functional derivatives with respect to these sources. We shall consider both the two-point and three-point correlation functions. The two-point correlation function is related to the power spectrum of the CMB anisotropy, while the three-point one to the nongaussianity of the spectrum.

To start with we consider the CTP path integral of the effective action $\Gamma$, including the influence action in Eq.~(\ref{SIFfinal}) as well as source terms for $\phi_{+}$ and $\phi_{-}$.
\begin{eqnarray}
&&e^{i\Gamma[J_+, J_-]} \nonumber \\
&=& \int D\phi_+ D\phi_- \int D\zeta P[\zeta] e^{i\int\left[-{1\over2}(\partial\phi_+)^2-{1\over2}V''(\Phi)\phi_+^2\right] +i\int J_+\phi_+}\nonumber\\
&&\hspace{5cm}e^{
-i\int\left[-{1\over2}(\partial\phi_-)^2-{1\over2}V''(\Phi)\phi_-^2\right] -i\int J_-\phi_-}  e^{-i\int \zeta\Delta\phi}  \nonumber \\
%&=&\int D\zeta P[\zeta] \int D\phi_+ D\phi_- e^{{i\over2}\int \phi_+ \left[\partial^2 -V''(\Phi)%%%\right]\phi_++ i\int (J_+-\zeta)\phi_+}
 %e^{-{i\over2}\int \phi_- \left[\partial^2 -V''(\Phi)\right]\phi_- - i\int (J_--\zeta)\phi_-}  \nonumber \\
&=& \int D\zeta P[\zeta] e^{-{i\over2}\int\left[(J_+-\zeta)G_{++}^\phi(J_+-\zeta) - (J_+-\zeta)G_{+-}^\phi(J_--\zeta)
- (J_--\zeta)G_{-+}^\phi(J_+-\zeta) + (J_--\zeta)G_{--}^\phi(J_--\zeta) \right]},\nonumber\\
\end{eqnarray}
where we have imposed the CTP boundary conditions to obtain the various Schwinger-Keldysh Green's functions. For the inflaton field these Green's functions are
\begin{eqnarray}
&&G_{++}^\phi (x,x') = -i \langle T\phi^{(0)}(x)\phi^{(0)}(x')\rangle, \quad G_{+-}^\phi (x,x') = -i \langle \phi^{(0)}(x')\phi^{(0)}(x)\rangle, \nonumber \\
&&G_{-+}^\phi (x,x') = -i \langle \phi^{(0)}(x)\phi^{(0)}(x')\rangle, \quad G_{--}^\phi (x,x') = -i \langle {\tilde T}\phi^{(0)}(x)\phi^{(0)}(x')\rangle,
\end{eqnarray}
where $\phi^{(0)}(x)$ is the free inflaton field, and $\tilde T$ is for anti-time ordered. 

The two-point correlator of the inflaton field, including the backreaction of the electromagnetic interaction, can be obtained by taking functional derivatives on the effective action with respect to the sources.
\begin{eqnarray}
&&\langle \phi(x)\phi(x')\rangle \nonumber \\
&=&\langle \phi_-(x)\phi_+(x')\rangle {\Big |}_{\phi_+=\phi_-=\phi} \nonumber \\
&=& \frac{1}{\sqrt{-g(x)}} \frac{\delta}{\delta J_-(x)}  \frac{1}{\sqrt{-g(x')}} \frac{\delta}{\delta J_+(x')} e^{i\Gamma[J_+, J_-]} {\Big |}_{J_+=J_-=0}  \nonumber \\
&=& \int D\zeta P[\zeta]\, \Big\{ i G_{-+}^\phi (x,x') - \int d^4x_1 \sqrt{-g(x_1)} \int d^4x_2 \sqrt{-g(x_2)} \,\times \nonumber \\
&&\left[G_{-+}^\phi (x,x_1) -G_{--}^\phi (x,x_1)\right]
\zeta(x_1)\zeta(x_2)  \left[-G_{++}^\phi (x_2,x') +G_{-+}^\phi (x_2,x')\right]+\cdots\Big\}.\label{2ptcorrelator}
\end{eqnarray}
Let us recall that the combination of the Green's functions
\begin{eqnarray}
G_{-+}^\phi (x,x_1) -G_{--}^\phi (x,x_1)&=&-i \langle \phi^{(0)}(x)\phi^{(0)}(x_1)\rangle + i \langle {\tilde T}\phi^{(0)}(x)\phi^{(0)}(x_1)\rangle \nonumber \\
% &=& -i \langle \phi(x)\phi(x_1)\rangle + i\theta(\tau-\tau_1) \langle \phi(x_1)\phi(x)\rangle +i\theta(\tau_1-\tau) \langle \phi(x)\phi(x_1)\rangle \nonumber \\
&=&- i\theta(\tau-\tau_1) \langle \left[\phi(x)^{(0)},\phi^{(0)}(x_1)\right]\rangle \nonumber \\
&=&-G_{\rm ret}^\phi(x,x_1), 
\end{eqnarray}
gives exactly the retarded Green function $G_{\rm ret}^\phi(x,x')$. The appearance of this retarded Green's function is related to the fact that the equation of motion derived from the in-in or Schwinger-Keldysh formalism respects causality explicitly. In a similar fashion, the other combination of the Green's functions also gives the retarded Green's function
\begin{eqnarray}
-G_{++}^\phi (x_2,x') +G_{-+}^\phi (x_2,x')&=&i \langle T\phi^{(0)}(x_2)\phi^{(0)}(x')\rangle - i \langle \phi^{(0)}(x_2)\phi^{(0)}(x')\rangle\nonumber\\
& =& G_{\rm ret}^\phi(x',x_2), 
\end{eqnarray}
Moreover, according to the definition of the probability density in Eq.~(\ref{N2}), the stochastic average of two $\zeta$'s is just the noise kernel $N_2(x_1,x_2)$. Therefore, the two-point correlator in Eq.~(\ref{2ptcorrelator}) becomes
\begin{eqnarray}
&&\langle \phi(x)\phi(x')\rangle\nonumber\\
&=& i G_{-+}^\phi (x,x') + \int d^4x_1 \sqrt{-g(x_1)} \int d^4x_2 \sqrt{-g(x_2)}  G_{\rm ret}^\phi(x,x_1) G_{\rm ret}^\phi(x',x_2) N_2(x_1,x_2) +\cdots\nonumber\\
\label{twopoint}
\end{eqnarray}
The first term is the usual two-point correlator of the inflaton vacuum fluctuations. Under quantization the inflaton field,
\begin{eqnarray}
\phi^{(0)}(x)=\int\,d^{3}k\,[a_{\vec{k}}\varphi_{\vec{k}}(\tau,\vec{x})+a^{\dagger}_{\vec{k}}\varphi_{\vec{k}}^{*}(\tau,\vec{x})]
\end{eqnarray}
where $\varphi_{\vec{k}}(\tau,\vec{x})$ is the mode function. During the inflationary epoch, the mode function can simply be represented by 
\begin{eqnarray}
\varphi_{\vec{k}}(\tau,\vec{x})=\left(-\frac{H}{2^{5/2}\pi\,k^{3/2}}\right)(-k\tau)^{3/2}H_{3/2}^{(1)}(-k\tau)\,e^{i\vec{k}\cdot\vec{x}}
\label{modefunction}
\end{eqnarray}
when the Bunch-Davies vacuum is chosen. With this mode function the first term above at equal conformal time can be readily evaluated to be
\begin{eqnarray}
\left.iG_{-+}^{\phi}\right|_{\tau'\rightarrow\tau}
&=&\int\frac{d^{3}k}{(2\pi)^{3/2}}\int\frac{d^{3}k'}{(2\pi)^{3/2}}e^{i\vec{k}\cdot\vec{x}}e^{i\vec{k}'\cdot\vec{x}'}\delta(\vec{k}+\vec{k}')\left(\frac{2\pi^{2}}{k^{3}}\right)P_{\phi}^{(0)},
\end{eqnarray}
where 
\begin{eqnarray}
P_{\phi}^{(0)}=\frac{H^{2}}{4\pi^{2}}(1+k^{2}\tau^{2})
\end{eqnarray}
is the power spectrum of the free inflaton field. 

The second term in Eq.~(\ref{twopoint}) gives the backreaction effect due to the electromagnetic interaction. It involves the retarded Green's function of the free inflaton field. In terms of the mode function in Eq.~(\ref{modefunction}), 
\begin{eqnarray}
G_{\rm ret}^{\phi}(x,x')&=&i\theta(\tau-\tau')\int d^{3}k\ \left[\varphi_{\vec{k}}(\tau,\vec{x})\varphi_{\vec{k}}^{*}(\tau',\vec{x}')-\varphi_{\vec{k}}^{*}(\tau,\vec{x})\varphi_{\vec{k}}(\tau',\vec{x}')\right]
\end{eqnarray}
Together with the noise kernel in Eq.~(\ref{noisekernel}), one can work on the integrals in this second term. As in the evaluation of the influence action in Section III, we shall again develop an asymptotic expansion in $1/\xi$. In this manner, at equal comformal time, the second term gives
\begin{eqnarray}
&&\left. \int d^4x_1 \sqrt{-g(x_1)} \int d^4x_2 \sqrt{-g(x_2)}\  G_{\rm ret}^\phi(x,x_1) G_{\rm ret}^\phi(x',x_2) N_2(x_1,x_2)\right|_{\tau'\rightarrow\tau}\nonumber\\
&=&\int\frac{d^{3}k}{(2\pi)^{3/2}}\int\frac{d^{3}k'}{(2\pi)^{3/2}}\,e^{i\vec{k}\cdot\vec{x}}e^{i\vec{k}'\cdot\vec{x}'}\delta(\vec{k}+\vec{k}')\left(\frac{\alpha^{2}}{f^{2}}e^{4\pi\xi}\right)\left(\frac{H^{4}}{4\pi^{2}k^{3}\xi^{8}}\right)\nonumber\\
&&\ \ \ \ \ e^{-4\sqrt{2\rho}}\left[\frac{105}{262144}\rho^{2}+\frac{15}{32768\sqrt{2}}\rho^{5/2}\right]+\dots
\end{eqnarray}
where we have defined $\rho\equiv-\xi k\tau$. From this result one can obtain the correction to the power spectrum \cite{barnaby},
\begin{eqnarray}
P_{\phi}^{(1)}=\left(\frac{\alpha^{2}}{f^{2}}\right)\left(\frac{H^{2}}{4\pi^{2}}\right)^{2}\left(\frac{e^{4\pi\xi}}{\xi^{8}}\right)\,h_{2}(\rho)
\end{eqnarray}
where the function
\begin{eqnarray}
h_{2}(\rho)=e^{-4\sqrt{2\rho}}\left[\frac{105}{131072}\rho^{2}+\frac{15}{16384\sqrt{2}}\rho^{5/2}\right]
\end{eqnarray}
The maximum value of this function 
\begin{eqnarray}
h_{2}|_{\rm max}\sim 5.86\times 10^{-6}
\end{eqnarray}
at $\rho\sim 0.6$.

Next we work on the three-point correlator. There are four types of functional derivatives:
\begin{eqnarray}
&&-\frac{1}{\sqrt{-g(x)}} \frac{\delta}{\delta J_+(x)}  \frac{1}{\sqrt{-g(x')}} \frac{\delta}{\delta J_+(x')} \frac{1}{\sqrt{-g(x'')}} \frac{\delta}{\delta J_+(x'')}
e^{i\Gamma[J_+, J_-]} {\Big |}_{J_+=J_-=0}  \nonumber \\
&=& i \langle T(\phi(x)\phi(x')\phi(x''))\rangle% \nonumber \\
%&=&  i\theta(\tau-\tau') \theta(\tau'-\tau'')\langle \phi(x)\phi(x')\phi(x'')\rangle +i\theta(\tau-\tau'') \theta(\tau''-\tau') \langle \phi(x)\phi(x'')\phi(x')\rangle \nonumber \\
%&& + i\theta(\tau'-\tau'') \theta(\tau''-\tau)\langle \phi(x')\phi(x'')\phi(x)\rangle +i\theta(\tau'-\tau) \theta(\tau-\tau'') \langle \phi(x')\phi(x)\phi(x'')\rangle \nonumber \\
%&& + i\theta(\tau''-\tau) \theta(\tau-\tau')\langle \phi(x'')\phi(x)\phi(x')\rangle +i\theta(\tau''-\tau') \theta(\tau'-\tau) \langle \phi(x'')\phi(x')\phi(x)\rangle,
\end{eqnarray}
\begin{eqnarray}
&&\frac{1}{\sqrt{-g(x)}} \frac{\delta}{\delta J_-(x)}  \frac{1}{\sqrt{-g(x')}} \frac{\delta}{\delta J_+(x')} \frac{1}{\sqrt{-g(x'')}} \frac{\delta}{\delta J_+(x'')}
e^{i\Gamma[J_+, J_-]} {\Big |}_{J_+=J_-=0}  \nonumber \\
&=& i \langle \phi(x)T(\phi(x')\phi(x''))\rangle %\nonumber \\
%&=&  i\theta(\tau'-\tau'') \langle \phi(x)\phi(x')\phi(x'')\rangle +i\theta(\tau''-\tau') \langle \phi(x)\phi(x'')\phi(x')\rangle,
\end{eqnarray}
\begin{eqnarray}
&&-\frac{1}{\sqrt{-g(x)}} \frac{\delta}{\delta J_-(x)}  \frac{1}{\sqrt{-g(x')}} \frac{\delta}{\delta J_-(x')} \frac{1}{\sqrt{-g(x'')}} \frac{\delta}{\delta J_+(x'')}
e^{i\Gamma[J_+, J_-]} {\Big |}_{J_+=J_-=0}  \nonumber \\
&=& i \langle {\tilde T}(\phi(x)\phi(x'))\phi(x'')\rangle %\nonumber \\
%&=&  i\theta(\tau-\tau') \langle \phi(x')\phi(x)\phi(x'')\rangle i\theta(\tau'-\tau) \langle \phi(x)\phi(x')\phi(x'')\rangle,
\end{eqnarray}
\begin{eqnarray}
&&\frac{1}{\sqrt{-g(x)}} \frac{\delta}{\delta J_-(x)}  \frac{1}{\sqrt{-g(x')}} \frac{\delta}{\delta J_-(x')} \frac{1}{\sqrt{-g(x'')}} \frac{\delta}{\delta J_-(x'')}
e^{i\Gamma[J_+, J_-]} {\Big |}_{J_+=J_-=0}  \nonumber \\
&=& i \langle {\tilde T}(\phi(x)\phi(x')\phi(x''))\rangle %\nonumber \\
%&=&  i\theta(\tau-\tau') \theta(\tau'-\tau'')\langle \phi(x'')\phi(x')\phi(x)\rangle +i\theta(\tau-\tau'') \theta(\tau''-\tau') \langle \phi(x')\phi(x'')\phi(x)\rangle \nonumber \\
%&& + i\theta(\tau'-\tau'') \theta(\tau''-\tau)\langle \phi(x)\phi(x'')\phi(x')\rangle +i\theta(\tau'-\tau) \theta(\tau-\tau'') \langle \phi(x'')\phi(x)\phi(x')\rangle \nonumber \\
%&& + i\theta(\tau''-\tau) \theta(\tau-\tau')\langle \phi(x')\phi(x)\phi(x'')\rangle +i\theta(\tau''-\tau') \theta(\tau'-\tau) \langle \phi(x)\phi(x')\phi(x'')\rangle.
\end{eqnarray}
It is obvious that each of them would converge to the same equal-time three-point correlation function
$\langle \phi(\tau,\vec{x})\phi(\tau,\vec{x}')\phi(\tau,\vec{x}''))\rangle$. For example, it is straightforward to show that
\begin{eqnarray}
&&\langle \phi(\tau,\vec{x})\phi(\tau,\vec{x}')\phi(\tau,\vec{x}''))\rangle \nonumber \\
&=& \langle T(\phi(x)\phi(x')\phi(x''))\rangle {\Big |}_{\tau>\tau'\rightarrow\tau>\tau''\rightarrow\tau} \nonumber \\
&=&i\frac{1}{\sqrt{-g(x)}} \frac{\delta}{\delta J_+(x)}  \frac{1}{\sqrt{-g(x')}} \frac{\delta}{\delta J_+(x')} \frac{1}{\sqrt{-g(x'')}} \frac{\delta}{\delta J_+(x'')}
e^{i\Gamma[J_+, J_-]} {\Big |}_{J_+=J_-=0,\,\tau>\tau'\rightarrow\tau>\tau''\rightarrow\tau} \nonumber \\
&=& -\int d^4x_1 \sqrt{-g(x_1)} \int d^4x_2 \sqrt{-g(x_2)} \int d^4x_3 \sqrt{-g(x_3)} \times \nonumber \\
&&\ \ \ \ \  G_{\rm ret}^\phi(x,x_1) G_{\rm ret}^\phi(x',x_2) G_{\rm ret}^\phi(x'',x_3) N_3(x_1,x_2,x_3) {\Big |}_{\tau>\tau'\rightarrow\tau>\tau''\rightarrow\tau},
\label{threepoint}
\end{eqnarray}
where $N_3$ is given by the stochastic force three-point function in Eq.~(\ref{Nthree}).

With the expression in Eq.~(\ref{threepoint}), we can now calculate the three point correlation function at equal conformal time as the two point correlator above. Again we are concerned with the asymptotic behavior of this three point correlator for large $\xi$. After evaluating the 6 terms due to permutations of $x_{1}$, $x_{2}$, and $x_{3}$ in $N_{3}(x_{1},x_{2},x_{3})$, we arrive at 
\begin{eqnarray}
&&\langle \phi(\tau,\vec{x})\phi(\tau,\vec{x}')\phi(\tau,\vec{x}''))\rangle \nonumber \\
&=&\int\frac{d^{3}k_{1}}{(2\pi)^{3/2}}\int\frac{d^{3}k_{2}}{(2\pi)^{3/2}}\int\frac{d^{3}k_{3}}{(2\pi)^{3/2}}\,e^{i\vec{k}_{1}\cdot\vec{x}}e^{i\vec{k}_{2}\cdot\vec{x}'}e^{i\vec{k}_{3}\cdot\vec{x}''}\delta(\vec{k}_{1}+\vec{k}_{2}+\vec{k}_{3})\left(\frac{1}{k^{6}}\right)\times\nonumber\\
&&\ \ \left(-\frac{3}{10}\right)(2\pi)^{5/2}\left(\frac{\alpha^{3}}{f^{3}}\right)\left(\frac{H^{2}}{4\pi^{2}}\right)^{3}\left(\frac{e^{6\pi\xi}}{\xi^{12}}\right)\left(\frac{1+x_{2}^{3}+x_{3}^{3}}{x_{2}^{3}x_{3}^{3}}\right)\,h_{3}(\rho;x_{2},x_{3})
\label{3ptcorrelator}
\end{eqnarray}
where we have parametrized the magnitudes of the vectors, $\vec{k}_{1}$, $\vec{k}_{2}$, and $\vec{k}_{3}$, constituting a triangle as $|\vec{k}_{1}|=k$, $|\vec{k}_{2}|=x_{2}k$, and $|\vec{k}_{3}|=x_{3}k$. The function $h_{3}(\rho;x_{2},x_{3})$ is given by
\begin{eqnarray}
h_{3}(\rho;x_{2},x_{3})
&=&\left(\frac{175}{25165824}\,\rho^{5}\right)\left[\frac{x_{2}^{3}x_{3}^{3}(1+x_{2}+x_{3})^{2}}{1+x_{2}^{3}+x_{3}^{3}}\right]\times\nonumber\\
&&\ \ \Bigg[\frac{(3+2\sqrt{x_{2}}+3x_{2})(1+x_{2}-x_{3})^{2}}{x_{2}^{5/2}(1+\sqrt{x_{2}})^{2}}e^{-4\sqrt{2\rho}(1+\sqrt{x_{2}})}\nonumber\\
&&\ \ \ \ +\frac{(3+2\sqrt{x_{3}}+3x_{3})(1-x_{2}+x_{3})^{2}}{x_{3}^{5/2}(1+\sqrt{x_{3}})^{2}}e^{-4\sqrt{2\rho}(1+\sqrt{x_{3}})}\nonumber\\
&&\ \ \ \ +\frac{(3x_{2}+2\sqrt{x_{2}}\sqrt{x_{3}}+3x_{3})(1-x_{2}-x_{3})^{2}}{x_{2}^{5/2}x_{3}^{5/2}(\sqrt{x_{2}}+\sqrt{x_{3}})^{2}}e^{-4\sqrt{2\rho}(\sqrt{x_{2}}+\sqrt{x_{3}})}\Bigg].
\end{eqnarray}
To get a feeling on the magnitude of $h_{3}$, we consider the case with the shape of an equilateral triangle where $x_{2}=x_{3}=1$. Here,
\begin{eqnarray}
h_{3}(\rho;1,1)=\left(\frac{525}{4194304}\rho^{5}\right)e^{-8\sqrt{2\rho}}.
\end{eqnarray}
The maximum value of this function 
\begin{eqnarray}
h_{3}|_{\rm max}\sim 1.65\times 10^{-9}
\end{eqnarray}
at $\rho\sim 0.78$. If we take $\rho=0.6$, where the maximum value of $h_{2}$ occurs, 
\begin{eqnarray}
h_{3}(0.6;1,1)=1.52\times 10^{-9}.
\end{eqnarray}
According to the way that Eq.~(\ref{3ptcorrelator}) is laid out, one can identify the nonlinear parameter of nongaussianity $f_{NL}$ of the inflaton field \cite{barnaby} as
\begin{eqnarray}
f_{NL}&=&-\left(\frac{\alpha^{3}}{f^{3}}\right)\left(\frac{e^{6\pi\xi}}{\xi^{12}}\right)\,h_{3}(\rho;x_{2},x_{3}).
\end{eqnarray}

\section{Conclusion}

We have studied the electromagnetic coupling effects in natural inflation in the Schwinger-Keldysh formalism.  The axion-photon-like coupling renders the spinoidal instability that leads to a copious production of helical electromagnetic fields during inflation. By tracing out the electromagnetic field, we have obtained the corresponding influence functional in the one-loop level, up to cubic order in the inflaton field. It is possible to interpret this functional as due to the effects of a single stochastic force. In this respect, the resulting probability density for the stochastic force will no long be gaussian. The two-point correlation function of the stochastic force is still given by the noise kernel. On the other hand, the three-point correlator will be nonvanishing. In turn, this will give rise to the nongaussianity effect of the inflaton field correlators.

We have also derived the equation of motin for the inflaton perturbation at one-loop level. It is in the form of a Langevin equation with the noise term originated from the fluctuations of the photon production. However, we have found that the dissipation term is absent in the Langevin equation. This seems to contradict the fluctuation-dissipation theorem. One should be reminded that the theorem holds only in the consideration of perturbative processes. It does not apply to the present case, because there is the spinoidal instability in which photon obtains a negative effective mass and becomes tachyonic, thus making all the Schwinger-Keldysh photon Green's functions equal. As a result, the photon production is a perfect energy sink rather than a dissipating process. 

The two-point and the three-point correlation functions are worked out explicitly here. It is interesting to note that only the retarded Green's function is involved since the in-in formalism is explicitly causal. The two-point correlator gives the correction to the usual power spectrum of the inflation field, whereas the three-point correlator gives the nongaussianity effect of the inflaton field. Both these results mostly agree with those in the previous work~\cite{anber,barnaby,meer}.

\begin{acknowledgments}
This work was supported in part by the Ministry of Science and Technology, Taiwan, ROC under the Grant Nos.~MOST104-2112-M-001-039-MY3 (K.W.N.), 
MOST107-2119-M-001-030 (K.W.N.), MOST106-2112-M-032-005 (H.T.C.), and MOST107-2112-M-032-011 (H.T.C.). HTC is also
supported in part by the National Center for Theoretical Sciences (NCTS). HTC would like
to thank the hospitality of the Theory Group of the Institute of Physics at the Academia
Sinica, Republic of China, where part of this work was done.
\end{acknowledgments}

\appendix

\section{Photon mode functions}
\label{mode}

Henceforth we assume that the spacetime is quasi de Sitter, meaning that $H$ and $d\Phi/dt$ are approximately constant. Therefore, we can treat $\xi$ as a constant parameter and that $a=e^{Ht}=-1/(H\tau)$. The photon mode equation can be obtained from the quadratic term in the action~(\ref{quadratic}) as
\begin{equation}
\left[\partial _{\tau}^{2} +k^2 \pm \frac{2\xi k}{\tau}\right] A_{R,L } (\vec{k},\tau )=0,
\end{equation}
where $k=|\vec{k}|$. To solve the mode equation, let us begin with the Whittaker equation:
\begin{equation}
\frac{d^2y}{dz^2}+\left[-{1\over 4}+{\kappa\over z}+\frac{1-\mu^2}{4z^2}\right]y=0.
\end{equation}
The independent solutions of this second-order differential equation  are the Whittaker functions, $W_{\kappa,\mu/2}(z)$ and $W_{-\kappa,\mu/2}(-z)$,
composed of the confluent hypergeometric functions~\cite{mathews}.
If we take $z=-2ik\tau$, $\kappa=\pm i\xi$, and $\mu=1$, then the photon mode equation will be the Whittaker equation with solutions,
\begin{equation}
A_{R,L}(k,\tau)=\frac{e^{\pm\xi \pi/2}}{\sqrt{2k}} W_{\mp i\xi,1/2} (2ik\tau),
\end{equation}
which are normalized such that as $k\tau\to-\infty$, they reduce to the adiabatic vacuum solutions, $A_{R,L}(k,\tau)\to e^{-ik\tau}/\sqrt{2k}$. Here in taking the limit,
we have used the asymptotic form of the Whittaker functions~\cite{buchholz}:
\begin{equation}
W_{\kappa,\mu/2} (z) \sim z^\kappa e^{-z/2} \left[1+O\left({1\over z}\right)\right],
\end{equation}
as $|z|\to\infty$ with $|{\rm arg} (z)|<3\pi/2$.
Next let us consider the behavior of the mode functions in limiting values of $\xi$. As $\xi\to 0$, they resume the plane wave solutions,
$A_{R,L}(k,\tau)= e^{-ik\tau}/\sqrt{2k}$. This is expected from the conformal invariance of a photon field in a conformally flat space-time that prohibits the growth of the photon field.
As $\xi\to \infty$, making use of the asymptotic forms of the Whittaker functions for large $\kappa$ with ${\rm Im} (\kappa)>0$ and ${\rm Im} (\kappa)<0$~\cite{buchholz}:
\begin{equation}
W_{\kappa,\mu/2} (z) \sim \left(\frac{z}{4\kappa}\right)^{1/4} e^{\kappa\ln\kappa-\kappa\mp i\left(\pi\kappa-\pi/4-2\sqrt{z\kappa}\right)},
\end{equation}
we obtain that
\begin{eqnarray}
A_{R}(k,\tau) &\sim& \frac{1}{\sqrt{2k}}\left(\frac{k|\tau|}{2\xi}\right)^{1/4} e^{\pi\xi-2\sqrt{2k|\tau|\xi}} e^{-i\xi\ln\xi+i\xi-i\pi/4},\label{modesol}\\
A_{L}(k,\tau) &\sim& \frac{1}{\sqrt{2k}}\left(\frac{k|\tau|}{2\xi}\right)^{1/4} e^{i\xi\ln\xi-i\xi+2i\sqrt{2k|\tau|\xi}}.
\end{eqnarray}
Note that $A_{R}(k,\tau)$ is amplified by a factor of $\exp (\pi\xi)$ as compared to $A_{L}(k,\tau)$. We thus conclude that the effect of photon particle production on inflation comes mainly from the $A_{R}$ modes. This justifies our treatment of neglecting the $A_{L}$ modes in the CTP calculation.


\begin{references}

\bibitem{guth}
A. H. Guth, Phys. Rev. D {\bf 23}, 347 (1981).

\bibitem{freese}
K. Freese, J. A. Frieman, and A. V. Olinto, Phys. Rev. Lett. {\bf 65}, 3233 (1990).

\bibitem{adams}
F. Adams, J. R. Bond, K. Freese, J. Frieman, and A. Olinto, Phys. Rev. D {\bf 47}, 426 (1993).

\bibitem{axion}
N. Arkani-Hamed, H.-C. Cheng, P. Creminelli, and L. Randall, Phys. Rev. Lett. {\bf 90}, 221302 (2003);
S. Dimopoulos, S. Kachru, J. McGreevy, and J. G. Wacker,  J. Cosmol. Astropart. Phys. 08 (2008) 003;
L. McAllister, E. Silverstein, and A. Westphal, Phys. Rev. D {\bf 82}, 046003 (2010);
N. Kaloper and L. Sorbo, Phys. Rev. Lett. {\bf 102}, 121301 (2009).

\bibitem{savage}
C. Savage, K. Freese, and W. H. Kinney, Phys. Rev. D {\bf 74}, 123511 (2006).

\bibitem{mohanty}
S. Mohanty and A. Nautiyal, Phys. Rev. D {\bf 78}, 123515 (2008).

\bibitem{anber}
M. M. Anber and L. Sorbo, Phys. Rev. D {\bf 81}, 043534 (2010).

\bibitem{visinelli}
L. Visinelli, J. Cosmol. Astropart. Phys. 09 (2011) 013.

\bibitem{mishra}
H. Mishra, S. Mohanty, and A. Nautiyal, Phys. Lett. B {\bf 710}, 245 (2012).

\bibitem{berera1}
A. Berera and L. Z. Fang,  Phys. Rev. Lett. {\bf 74}, 1912 (1995).

\bibitem{berera2}
A. Berera,  Phys. Rev. Lett. {\bf 75}, 3218 (1995).

\bibitem{wolung}
W. Lee and L. Z. Fang,  Int. J. Mod. Phys. D {\bf 6}, 305 (1997); Phys. Rev. D {\bf 59}, 083503 (1999).

\bibitem{berera3}
A. Berera, I. G. Moss, and R. O. Ramos, Rept. Prog. Phys. {\bf 72}, 026901 (2009).

\bibitem{anber2}
M. M. Anber and L. Sorbo, J. Cosmol. Astropart. Phys. 10 (2006) 018.

\bibitem{durrer}
R. Durrer, L. Hollenstein, and R. K. Jain, J. Cosmol. Astropart. Phys. 03 (2011) 037.

\bibitem{barnaby}
N. Barnaby and M. Peloso, Phys. Rev. Lett. {\bf 106}, 181301 (2011);
N. Barnaby, R. Namba, and M. Peloso, J. Cosmol. Astropart. Phys. 04 (2011) 009.

\bibitem{meer}
P. D. Meerburg and E. Pajer, J. Cosmol. Astropart. Phys. 02 (2013) 017.

\bibitem{ng}
K.-W. Ng, Phys. Rev. D {\bf 86}, 103510 (2012).

\bibitem{barnaby2}
N. Barnaby, E. Pajer, and M. Peloso, Phys. Rev. D {\bf 85}, 023525 (2012).

\bibitem{cook}
J. L. Cook and L. Sorbo, Phys. Rev. D {\bf 85}, 023534 (2012);
{\bf 86}, 069901(E) (2012);
M. M. Anber and L. Sorbo, Phys. Rev. D {\bf 85}, 123537 (2012);
N. Barnaby, J. Moxon, R. Namba, M. Peloso, G. Shiu, and P. Zhou, Phys. Rev. D {\bf  86}, 103508 (2012).

\bibitem{lin}
C.-M. Lin and K.-W. Ng, Phys. Lett. B {\bf 718}, 1181 (2013).

\bibitem{linde}
A. Linde, S. Mooij, and E. Pajer, Phys. Rev. D {\bf 87}, 103506 (2013).

\bibitem{bugaev}
E. Bugaev and P. Klimai, Phys. Rev. D {\bf 90}, 103501 (2014).

\bibitem{fugita}
T. Fujita, R. Namba, Y. Tada, N. Takeda, and H. Tashiro,
J. Cosmol. Astropart. Phys. 05 (2015) 054.

\bibitem{adshead}
P. Adshead, J. T. Giblin, T. R. Scully, and E. I. Sfakianakis, J. Cosmol. Astropart. Phys. 12 (2015) 034.

\bibitem{cheng}
S.-L. Cheng, W. Lee, and K.-W. Ng, Phys. Rev. D {\bf 93}, 063510 (2016).

\bibitem{ferreira}
R. Z. Ferreira, J. Ganc, J. Nore\~{n}a, and M. S. Sloth,  J. Cosmol. Astropart. Phys. 04 (2016) 039;
 J. Cosmol. Astropart. Phys. 10 (2016) E01.

\bibitem{jordan}
R. D. Jordan, Phys. Rev. D {\bf 33}, 44 (1986).

\bibitem{calzetta}
E. Calzetta and B. L. Hu, Phys. Rev. D {\bf 35}, 495 (1987).

\bibitem{schwinger}
J. S. Schwinger, J. Math. Phys. {\bf 2}, 407 (1961).

\bibitem{keldysh}
L. V. Keldysh, Zh. Eksp. Teor. Fiz. {\bf 47}, 1515 (1964).

\bibitem{feynman}
R. Feynman and F. Vernon, Ann. Phys. (N.Y.) {\bf 24} 118 (1963).


\bibitem{calzettahu}
E. Calzetta and B. L. Hu, {\em Nonequilibrium Quantum Field Theory}, Cambridge University Press (2008).

\bibitem{benderorszag}
C. M. Bender and S. A. Orszag, {\em Advanced Mathematical Methods for Scientists and Engineers}, McGraw-Hill (1978).


\bibitem{mathews}
J. Mathews and R. L. Walker, {\em Mathematical Methods of Physics (2nd Edition)}, Addison-Wesley (1971).

\bibitem{buchholz}
H. Buchholz, {\em The Confluent Hypergeometric Function}, Springer (1969).

\end{references}
\end{document}